\documentclass[letterpaper, 10 pt, conference]{ieeeconf}  
\IEEEoverridecommandlockouts                              
\usepackage{epsfig} 
\usepackage{graphicx}
\usepackage{amsmath}
\usepackage{array}
\usepackage{amssymb}
\usepackage{verbatim}
\usepackage{cleveref}
\usepackage{color}
\usepackage{relsize}
\usepackage{fancyhdr}
\usepackage{graphics}
\usepackage{setspace}
\usepackage{epstopdf}
\usepackage{multirow}
\usepackage{lipsum}
\usepackage{textcomp}
\usepackage{mathtools}
\usepackage{cite}
\usepackage{setspace}
\usepackage{booktabs}

\fancypagestyle{firstpage}{\lhead{\vspace{-0.75 cm} \small \textbf{{\fontfamily{cmss}\selectfont
The 3rd IEEE Conference on Control Technology and Applications (CCTA)\\August 19--21, 2019, Hong Kong, China}}}}

\title{\LARGE \bf
MPC-based Precision Cooling Strategy (PCS) for Efficient Thermal Management of Automotive Air Conditioning System}

\author{Hao Wang$^{1}$, Yan Meng$^{2}$, Quansheng Zhang$^{2}$, Mohammad Reza Amini$^{1}$, Ilya Kolmanovsky$^{3}$,\\ Jing Sun$^{1}$, and Mark Jennings$^{2}$
\thanks{$^{1}$Hao Wang, Mohammad Reza Amini, and Jing Sun are with the Department of Naval Architecture \& Marine Engineering, University of Michigan, Ann Arbor, MI 48109 USA. Emails: {\tt\small \{autowang, mamini, jingsun\}@umich.edu}}%
\thanks{$^{2}$Yan Meng, Quansheng Zhang, and Mark Jennings are with Ford Motor Company, Dearborn, MI 48124 USA. Emails: {\tt\small \{ymeng, qzhang71, mjennin5\}@ford.com}}
\thanks{$^{3}$Ilya Kolmanovsky is with  the Department of Aerospace Engineering, University of Michigan, Ann Arbor, MI 48109 USA. Email: {\tt\small ilya@umich.edu}}%
}

\renewcommand{\baselinestretch}{1}
\begin{document}

\maketitle
\thispagestyle{firstpage}

\begin{abstract}
In this paper, we propose an MPC-based precision cooling strategy (PCS) for energy efficient thermal management of automotive air conditioning (A/C) system. The proposed PCS is able to provide precise tracking of the time-varying cooling power trajectory, which is assumed to match the passenger comfort requirements. In addition, by leveraging the emerging connected and automated vehicles (CAVs) technology, vehicle speed preview can be incorporated in our A/C thermal management strategy for further energy efficiency improvement. This proposed A/C thermal management strategy is developed and evaluated based on a physics-based A/C system model (ACSim) from Ford Motor Company for the vehicles with electrified powertrains. In a comparison with Ford benchmark case over SC03 cycle, for tracking the same cooling power trajectory, the proposed PCS provides $4.9\%$ energy saving at the cost of slight increase in the cabin temperature (less than $1$$^oC$). It is also demonstrated that by coordinating with future vehicle speed and shifting the A/C power load, the A/C energy consumption can be further reduced.    


\end{abstract}

\section{INTRODUCTION}

Emerging connected and automated vehicles (CAVs) technologies are pushing vehicle safety and energy efficiency to the next level and create unprecedented opportunities and challenges for the control and optimization of the vehicle systems. While previous studies have been focusing on improving the fuel efficiency via powertrain optimizations (\hspace{-0.01cm}\cite{Zhang2011}, \cite{Zhang2012}, and \cite{Guanetti18}), vehicle thermal management and its interaction with powertrain control in hot and cold weather conditions have not been fully explored.

Typical thermal management systems in ground vehicles include the engine cooling \cite{Amey2016}, exhaust heat recovery \cite{Feru2016}, battery and electric machine thermal management \cite{Xia2017} \cite{Snyder2014}, and cabin heating, ventilation, \& air conditioning (HVAC) system. In light-duty passenger cars, the power consumed by the HVAC system, which creates a comfortable passenger compartment, represents the most significant auxiliary load\cite{Rugh2008}, which can significantly impact the overall vehicle fuel efficiency. In the United States, an estimated 7 billion gallons of fuel is consumed annually for the air conditioning (A/C) system of light-duty vehicles~\cite{Rugh2008}. Over 50\% range reduction due to heating the cabin in winter has been observed in EV tests for the UDDS driving cycle, which were performed at Argonne National Lab \cite{Jeffers2016}. Range reduction due to A/C operation in the summer time is slightly lower for EV as reported in \cite{Rask2014}, and \cite{Jeffers2015}, however, the impact of A/C operation on vehicle energy consumption is still considerable and the comparison with the heating counterpart can be quite different for different driving cycles and powertrain configurations. For example, regarding the HEV applications, since the cabin heating may utilize the engine coolant heat, its impact on fuel economy may not be as dramatic as the one in the EV applications. 

Aiming at reducing vehicle-level fuel consumption, our previous efforts have been focused on developing energy efficient thermal management strategies for the A/C system. In \cite{Zhang2017}, the analysis of A/C system was performed and the optimal compressor and fan speed controls have been investigated. We also studied the speed sensitivity of the A/C system efficiency in\cite{Hao2018}, which has been exploited to reduce the A/C system energy consumption via model predictive control (MPC). Reference \cite{Reza18} demonstrated the impact of uncertain traffic information on optimizing the A/C energy efficiency and evaluated the overall vehicle fuel economy over different driving cycles. 

In this paper, a precision cooling strategy (PCS) is proposed, attempting to address the trade-offs between the occupant thermal comfort and the A/C system energy consumption \cite{Zhang2017}. In order to quantify such trade-offs, a new performance metric, discharge air cooling power (DACP), is defined as follows:
\begin{eqnarray}
\label{eqn:DACP}
P_{DACP}(t)=c_p(T_{cab}(t)-T_{discharge}(t))W_{bl}(t),
\end{eqnarray} 
where $c_p$ is the specific heat capacity of air, $T_{cab}$ represents the average cabin temperature, $T_{discharge}$ represents the discharge air temperature, namely, the temperature of the air after the heat exchange with the evaporator, and $W_{bl}$ represents the air flow rate into the cabin delivered by HVAC blower. Note that the DACP in (\ref{eqn:DACP}) is defined for the case when A/C is running in the recirculation mode, which is also the simulation condition investigated hereafter. If fresh air mode is considered, $T_{cab}$ should be replaced by $T_{amb}$ (ambient temperature). The integral of DACP over time is referred to as the discharge air cooling energy (DACE) and it is denoted by $E_{DACE}$. A key assumption behind this definition is that there exists a time-varying trajectory of $P_{DACP,targ}$ that, if it is precisely tracked, the occupant comfort requirement can be satisfied. In the definition, two major variables, temperature and flow rate of the cooling air, are considered to primarily impact the comfort. Compared with the average room temperature, which is commonly used as the performance metric in building HVAC control \cite{Kelman11}, \cite{Oldewurtel12} and also in our previous works \cite{Hao2018}, \cite{Reza18, Reza19_2}, the choice of $P_{DACP}$ accounts for special characteristics of the automotive A/C system. In a passenger vehicle, occupants sit close to the vents and directly feel the temperature and the amount of air flow. The occupants' sensation to A/C is therefore not directly correlated to average cabin temperature but instead may be better captured by the new performance index proposed here. We note that realistic occupant comfort requirements are much more complicated than the $P_{DACP,targ}$ metric defined here and that research is currently ongoing to define better performance metrics for guiding the design of HVAC control systems in automotive applications. Besides the precise tracking of $P_{DACP,targ}$ which is intended to prevent over-cooling of the cabin, the idea similar to \cite{Hao2018} of exploiting the speed sensitivity of A/C system efficiency will also be pursued.  

Specifically, the work presented in this paper may be directly compared with our previous work done in \cite{Hao2018} since they both explore the speed sensitivity of the A/C system and apply the model predictive control (MPC) design framework. However, the proposed MPC-based PCS has the following features that differentiate it from that of \cite{Hao2018}:
\begin{enumerate}
	\item A new performance metric, DACP, is proposed instead of applying average cabin temperature in \cite{Hao2018} and \cite{Reza18}. The proposed PCS is designed to precisely track the prescribed and possibly time-varying trajectory of DACP. 
	\item A new predictive model structure is proposed which, unlike the one used in \cite{Hao2018}, takes into account the transient effect of air flow rate on evaporator wall temperature. 
	\item The proposed PCS coordinates the A/C operation with vehicle speed by simply manipulating the design parameters in the cost function of the MPC problem. Note that in our previous work, such a coordination was achieved by manipulating the operating constraints, which requires additional design efforts.  
\end{enumerate}\vspace{-0.1cm} 

The rest of the paper is organized as follows. Section~\ref{sec:1} introduces the A/C system in a power-split HEV and the physics-based system model. Next, the predictive model development is described in Section~\ref{sec:2}. The design procedures of the PCS are detailed in Section~\ref{sec:3}. Section~\ref{sec:4} presents the simulation results of the proposed strategy, which demonstrates energy saving potentials, followed by the conclusions in Section~\ref{sec:5}.

\section{System Description and High-fidelity ACSim Model}\label{sec:1}
\subsection{A/C System in Power-split HEVs}
\begin{figure}[h!]
	\begin{center}
		\includegraphics[width=8.5cm]{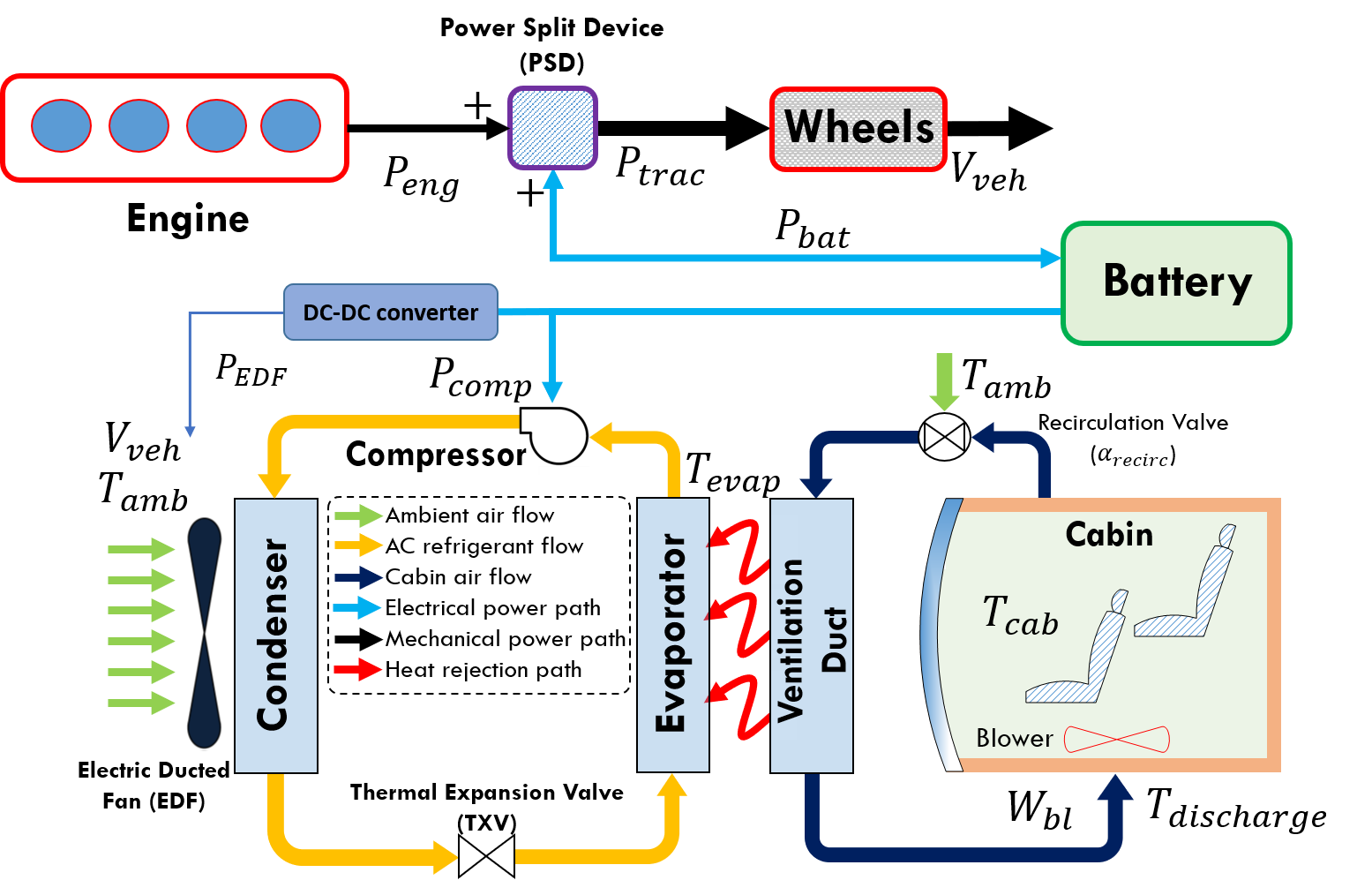} \vspace{-0.2cm}   
		\caption{Schematic of the A/C system in a power-split HEV}\vspace{-0.5cm}
		\label{fig:iPTM_system}
	\end{center}
\end{figure}

A typical A/C system configuration for power-split HEVs is considered in this paper, see Fig.~\ref{fig:iPTM_system}. There are two major loops within the A/C system, the vapor compression loop shown in yellow and the air supply loop shown in dark blue. In this HEV configuration, battery directly supplies the electrical power for the electrically driven compressor and the electric ducted fan (EDF), which consume most of the energy in the A/C system. Assuming a charge sustaining operation for the battery, the energy consumed by the A/C system will be eventually supplied from the fuel energy, converted by the engine and the power split device (PSD). 
\vspace{-0.2cm}

\subsection{ACSim Model and Speed Sensitivity Analysis}
\begin{figure}[h!]
	\begin{center}
		\includegraphics[width=7.5cm]{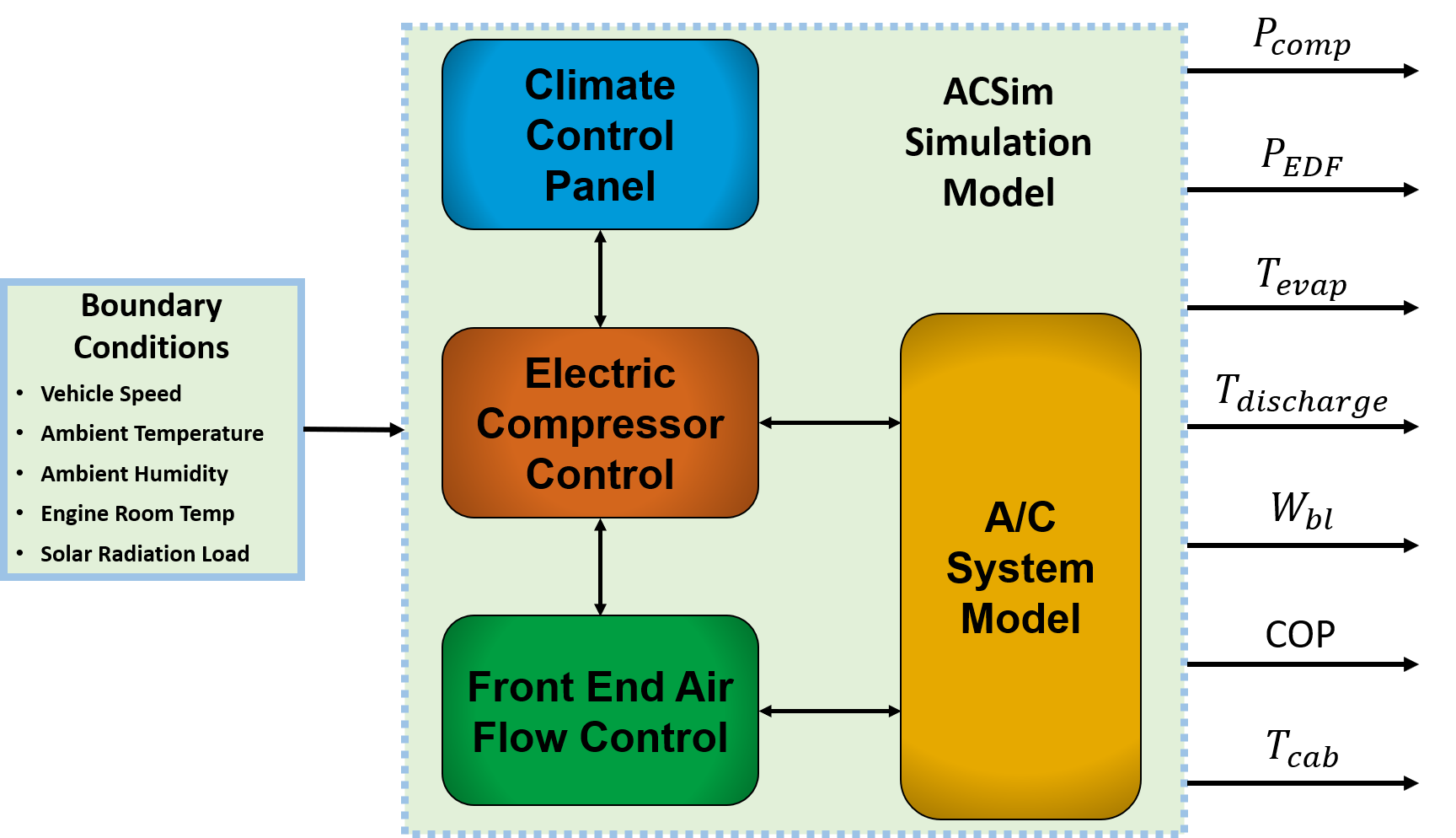} \vspace{-0.2cm}   
		\caption{Schematics of ACSim simulation model}\vspace{-0.5cm}
		\label{fig:ACSIM}
	\end{center}
\end{figure}

Physics-based modeling of the A/C system can be very challenging \cite{Zhang16}, especially for modeling the vapor compression cycle shown in Fig.~\ref{fig:iPTM_system}. In this study, we utilize the Ford A/C system model, which is referred to as ACSim, for control design and validation purposes. General system schematics are illustrated in Fig.~\ref{fig:ACSIM}. This model simulates the entire A/C system for a passenger car and is integrated with the controller module which represents two levels of controls. A higher-level controller is inside the climate control panel block, and it reflects the control settings (e.g. blower level and temperature set-point) from the real vehicle, which directly affect the occupant thermal comfort. Lower-level controllers take the command from the control panel and regulate the behaviors of the physical system via the electric compressor control and the front end air flow control. Boundary conditions are set according to different simulation requirements.
\vspace{-0.25cm}

\begin{figure}[h!]
	\begin{center}
		\includegraphics[width=7cm]{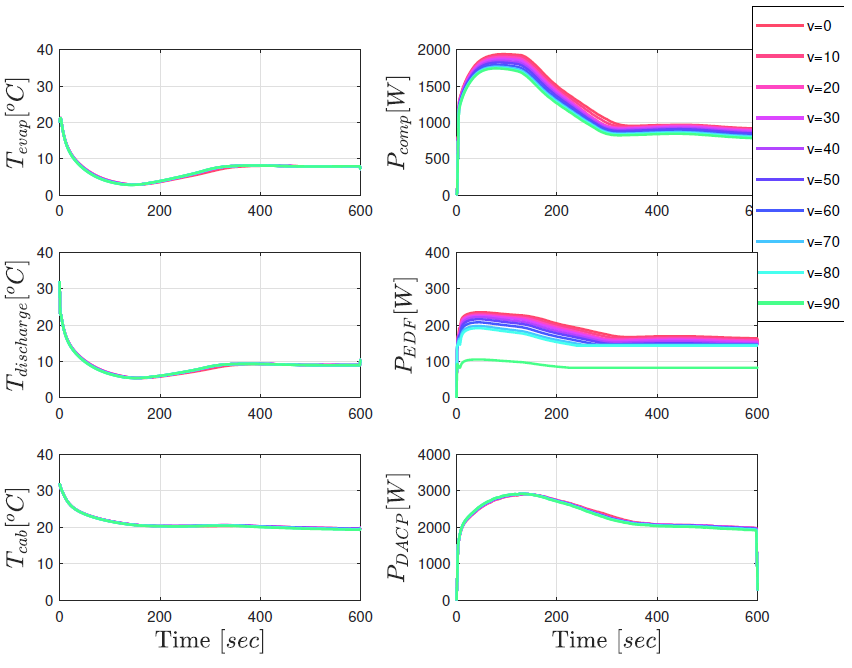} \vspace{-0.5cm}   
		\caption{Sensitivity of the ACSim model responses to vehicle speed.} \vspace{-0.5cm}
		\label{fig:SpeedSensitivity1} 
	\end{center}\vspace{-0.35cm}
\end{figure}

\begin{figure}[h!]
	\begin{center}
		\includegraphics[width=4.5cm]{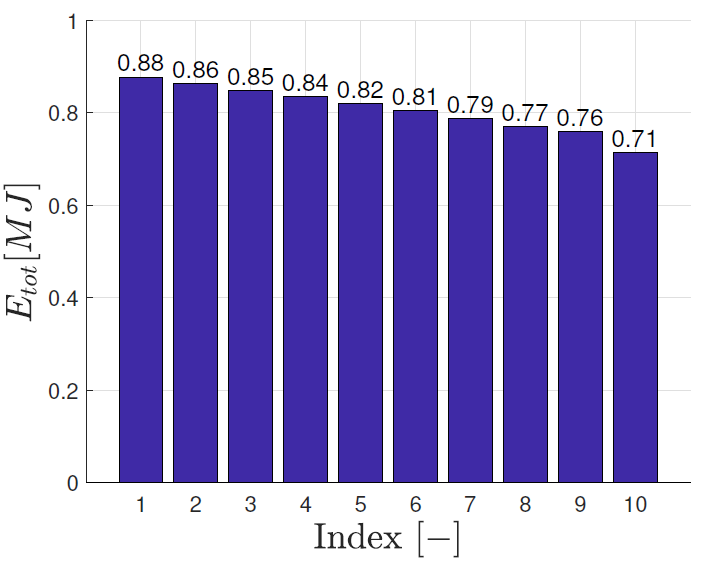} \vspace{-0.3cm}   
		\caption{Total A/C energy consumption decreases as vehicle speed increases.} \vspace{-0.55cm}
		\label{fig:SpeedSensitivity2} 
	\end{center}
\end{figure}

Next, the speed sensitivity of A/C system energy consumption is demonstrated. In Fig.~\ref{fig:SpeedSensitivity1}, it can be seen that with almost the same $P_{DACP}$ trajectories (instantaneous deviation within $1\%$), the power trajectories for compressor and EDF shift downwards as vehicle speed increases. Fig.~\ref{fig:SpeedSensitivity2} summarizes the total A/C energy consumption ($E_{tot}=E_{comp}+E_{EDF}$, where $ E_{comp}$ and $E_{EDF}$ represent the energy consumed by compressor and EDF, respectively) for each case shown in Fig.~\ref{fig:SpeedSensitivity1}. Index values from $1$ to $10$ correspond to constant vehicle speed values from $0$ $km/h$ to $90$ $km/h$, respectively. As the simulation results show, the total A/C energy consumption is reduced by $13.6\%$ comparing case $10$ with case $1$, while the cooling performance is kept the same. This observation is consistent with the findings in \cite{Hao2018}. This speed sensitivity of A/C system efficiency will be exploited in the PCS design for reducing the energy consumption.
\vspace{-0.4cm}

\vspace{-0.2cm}

\section{Simplified A/C System Model for MPC design}\label{sec:2}
\subsection{Predictive Model Structure}
Like other high-fidelity A/C system models \cite{Zhang16}, ACSim model involves detailed thermal and fluid dynamics of the refrigerant and has a large number of look-up tables from calibrations, which make it impossible to be used in a controller design. Therefore, a simplified model of the system dynamics is necessary. Specifically, the following discrete-time model structure is proposed to satisfy the requirements for MPC-based design: 
%
\begin{eqnarray}
\label{eqn:1}
T_{evap}(k+1) &=&f_{T_{evap}}=T_{evap}(k) \nonumber\\ &+&\gamma_1(T_{evap}(k)-T_{evap,targ}(k)) \nonumber\\ &+&\gamma_2(T_{evap}(k)-T_{amb})W_{bl}(k) \nonumber\\ &+&\gamma_{3}(T_{evap}(k)-T_{amb})\Delta W_{bl}(k)+\gamma_4, \hspace{+0.5cm}\\
\label{eqn:2}
W_{bl}(k+1)&=&f_{W_{bl}}= W_{bl}(k)+\Delta W_{bl}(k),\\
\label{eqn:3}
T_{discharge}(k)&=&f_{T_{discharge}}\nonumber\\
&=&\gamma_5T_{evap}(k)+\gamma_6T_{cab}(k)+\gamma_7.
\end{eqnarray} 
In (\ref{eqn:1})-(\ref{eqn:3}), $T_{cab}$, $T_{evap}$, $T_{amb}$, $W_{bl}$ and $T_{discharge}$ represent the cabin average air temperature, the evaporator wall temperature, the ambient temperature, the blower air flow rate, and the discharge air temperature, respectively. All temperatures are in $^oC$ and the blower air flow rate has the units of $kg/s$. The model states are $T_{evap}$ and $W_{bl}$. The model inputs are the incremental blower air flow rate, $\Delta W_{bl}$, and the evaporator wall temperature target, $T_{evap,targ}$. The model parameters, $\gamma_i~(i=1,2,...,7)$, are constants and to be identified for matching the system responses. This predictive model is nonlinear because of the multiplicative coupling between model states and inputs in (\ref{eqn:1}). 


Compared with the evaporator wall temperature model proposed in \cite{Hao2018}, which is modeled as a first-order system with $T_{evap,targ}$ as an input, air flow effects ($W_{bl}$ and $\Delta W_{bl}$) are considered in this work based on the observation that with fixed $T_{evap,targ}$, $T_{evap}$ changes when air flow changes. 

\subsection{Model Identification and Validation}
Next, the ACSim model is simulated with different random sinusoidal input signals. The system responses are collected with the sampling time, $T_{s}=3 sec$, to identify the unknown parameters in (\ref{eqn:1}) and (\ref{eqn:3}). The resulting identified parameters are $\gamma=\left[\gamma_1~\gamma_2~ ... ~\gamma_7\right]=\left[-0.084,-0.487,-1.121,-1.730,0.729,0.690,-11.457\right]$.

\begin{figure}[h!]
	\begin{center}
		\includegraphics[width=7cm]{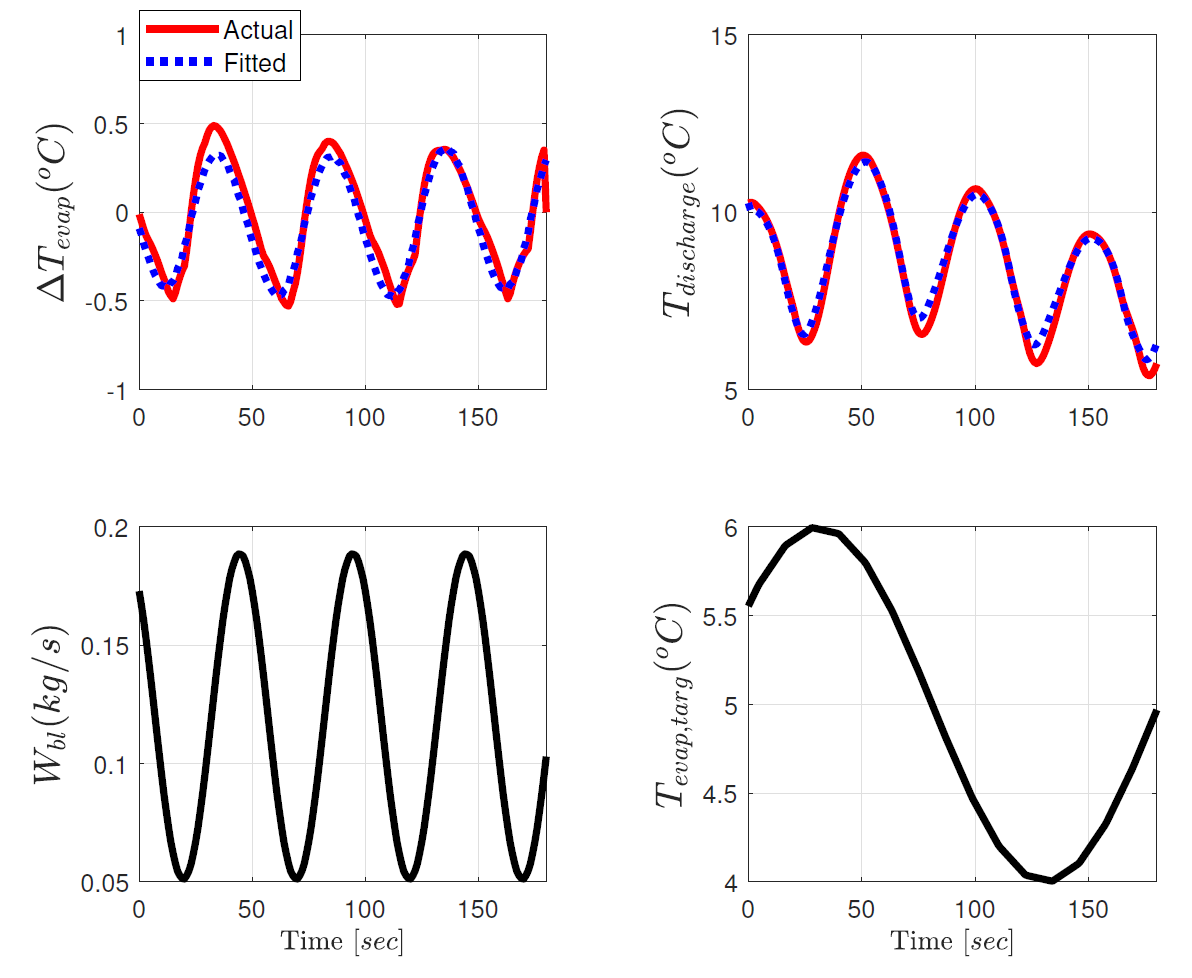} \vspace{-0.45cm}     
		\caption{Model validation results of $\Delta T_{evap}(k)=T_{evap}(k+1)-T_{evap}(k)$ and $T_{discharge}(k)$ for given sinusoidal excitations.}     
		\label{fig:Prediction1}
	\end{center} 
\end{figure}

Fig.~\ref{fig:Prediction1} provides the validation results of the simplified predictive model for matching the outputs from ACSim model. It confirms the good accuracy of the proposed model in modeling the key dynamics of the A/C system.
\vspace{-0.3cm}
\vspace{-0.65cm}

\section{MPC-based Precision Cooling Strategy (PCS)}\label{sec:3} 
\begin{figure}[h!]
	\begin{center}
		\includegraphics[width=5.0cm]{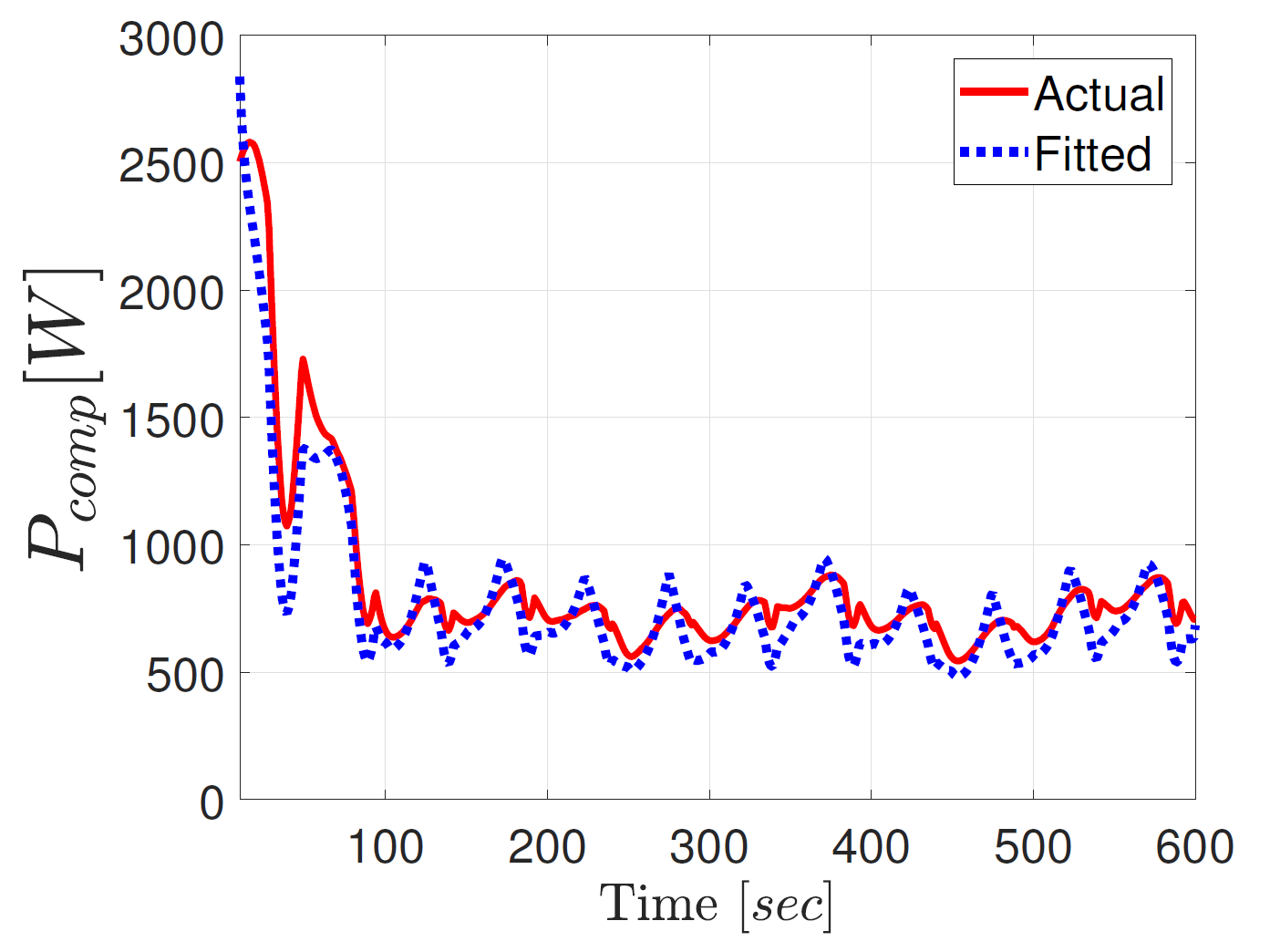} \vspace{-0.3cm}     
		\caption{Estimated compressor power based on (\ref{eqn:Pc}) compared with actual compressor power measured from ACSim.}    
		\label{fig:Pcomp_est}
	\end{center} 
\end{figure}

In this section, the problem formulation of the proposed PCS is described, whose objective is combining the minimization of overall A/C energy consumption and the tracking error with respect to the target $P_{DACP,targ}$ trajectory. As may be observed in Fig.~(\ref{fig:SpeedSensitivity1}), the compressor power is dominant as compared with the EDF power. Therefore, we decide to use the predicted compressor power in the cost function to reflect the overall system energy consumption in the proposed nonlinear MPC (NMPC) problem. According to \cite{Kelman11}, $P_{comp}$ can be estimated by:
\begin{eqnarray}
\label{eqn:Pc}
P_{comp}=\frac{c_p}{COP(k)}(T_{cab}(k)-T_{discharge}(k))W_{bl}(k), 
\end{eqnarray}
where $c_p=1008~J/(kg\cdot K)$ is the specific heat capacity of air at constant pressure, $COP$ represents the A/C system coefficient of performance \cite{Bhatti97}. Note that, COP may be time-varying, however, in the MPC problem formulation, it is assumed to be constant over the prediction horizon and will be updated based on current measurements at the beginning of each control iteration. Fig.~\ref{fig:Pcomp_est} shows the comparison between the compressor power estimated using (\ref{eqn:Pc}) and the actual compressor power computed by ACSim, which is based on the thermo-dynamics of the vapor-compression refrigeration system.
\vspace{-0.5cm} 
\vspace{-0.5cm}

Then, we define the PCS strategy as the following nonlinear optimization problem: 
\begin{equation} 
\begin{aligned}\label{eqn:MPC-PCS}
&\min_{\substack{\Delta W_{bl}\\ T_{evap,targ}}} && \sum_{i=0}^{N_p} \Bigg\{ \begin{gathered} P_{comp}(i|k)+\alpha\cdot(P_{DACP}(i|k)\\-\beta(i|k)\cdot P_{DACP,targ}(i|k))^2 \end{gathered}\Bigg\},\\
& \text{s.t.}
& & T_{evap}(i+1|k)=f_{T_{evap}}(i|k),\\
&
& & W_{bl}(i+1|k)=f_{W_{bl}}(i|k),\\
&
& &0~^oC\leq T_{evap}(i|k)\leq \overline{T}_{evap}(i|k),\\
&
& &0.05~kg/s \leq W_{bl}(i|k)\leq 0.15~kg/s,\\
&
& &-0.05~kg/s \leq \Delta W_{bl}(i|k)\leq 0.05~kg/s,\\
&
& &2~^oC\leq T_{evap,targ}(i|k)\leq 10~^oC,\\
&
& & T_{evap}(0|k)=T_{evap}(k),~W_{bl}(0|k)=W_{bl}(k).
\end{aligned}
\end{equation}

In (\ref{eqn:MPC-PCS}), $(i|k)$ denotes the prediction for the time instant $k+i$ made at the time instant $k$, $f_{T_{evap}}$ and $f_{W_{bl}}$ are from (\ref{eqn:1}) and (\ref{eqn:2}). In the cost function, $\alpha$ and $\beta$ are design parameters. In this study, $\alpha$ is set to be a large positive constant, e.g., $10^5$, to ensure the tracking performance. While $\beta$ can be either constant, $1$, or time-varying with respect to vehicle speed preview, depending on the operating scenarios of the A/C system. Detailed design of $\beta$ and its impact will be discussed in the next section. $P_{DACP,targ}$ and $\overline{T}_{evap}$ represent the target DACP trajectory and the time-varying upper bound for $T_{evap}$, respectively, which are assumed to be known over the prediction horizon. Constant constraints for other variables are given according to the system operating requirements. For the results presented in the next section, the prediction horizon, $N_p$, is set to be $10$. The NMPC problem described by (\ref{eqn:MPC-PCS}) is solved numerically using the MPCTools package \cite{mpctools}. This package exploits CasADi \cite{CAS18} for automatic differentiation and IPOPT algorithm for the numerical optimization. 

\section {Simulation Results and Performance Evaluations} \label{sec:4}
\subsection{Simulation Results on the Simplified Model}
\begin{figure}[h!]
	\begin{center}
		\includegraphics[width=7.5cm]{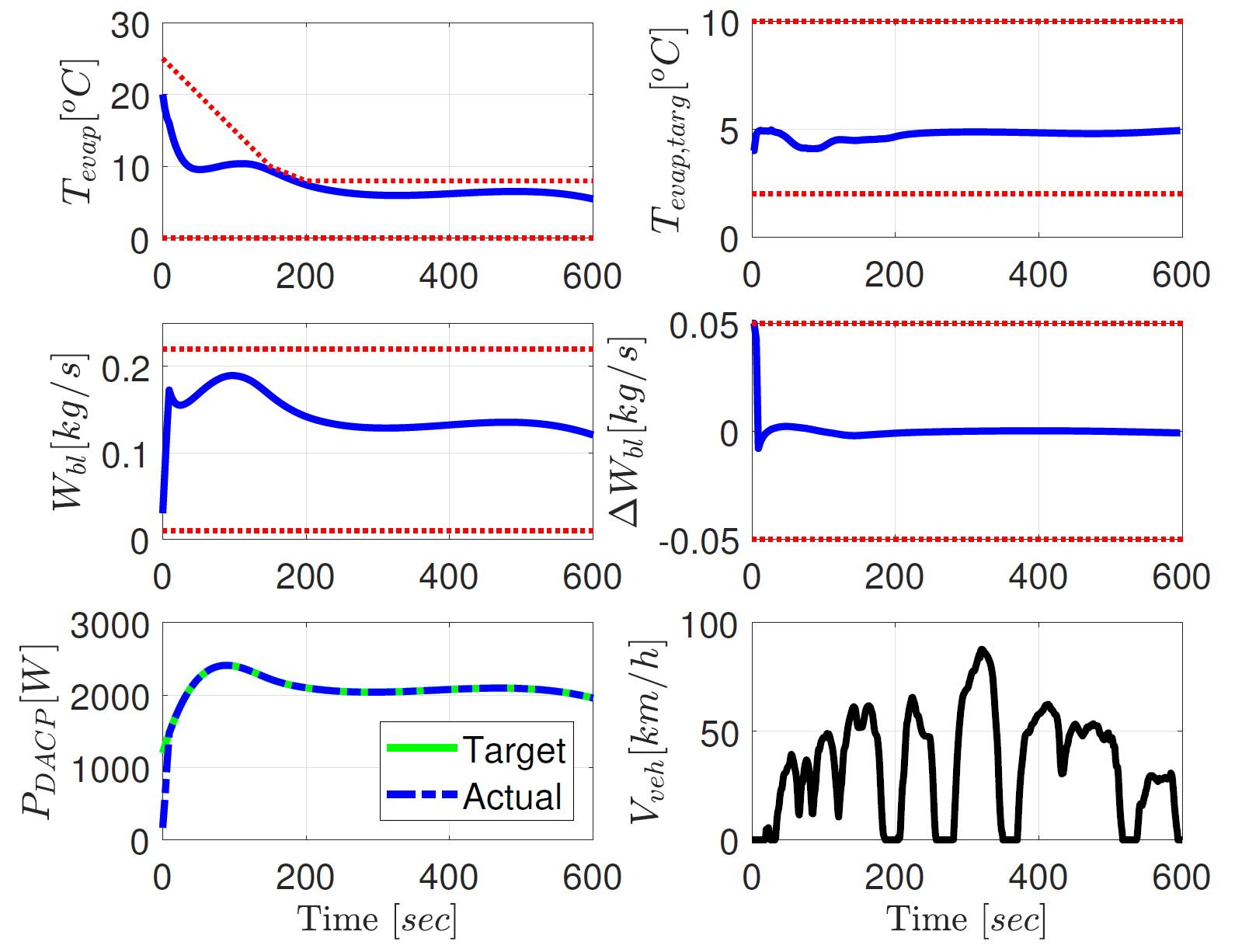}\vspace{-0.35cm}   
		\caption{Performance evaluation of the proposed PCS on the simplified A/C system model.} 
		\label{fig:MPC_SimpleModel} \vspace{-0.65cm}
	\end{center}
\end{figure}

The performance of the proposed MPC-based PCS is first evaluated on the simplified system model developed in Section~\ref{sec:2}. In Fig.~\ref{fig:MPC_SimpleModel}, an example of simulating a typical summer cabin cool-down scenario is shown. In order to ensure precise tracking of $P_{DACP,targ}$, constant $\beta=1$ is set. Vehicle speed trajectory from SC03 cycle is applied. It can been from Fig.~\ref{fig:MPC_SimpleModel} that all the state and input constraints in red dotted lines are satisfied, and perfect tracking of $P_{DACP,targ}$ is achieved except for the initial transient period. In this simulation, $T_{cab}$ and $COP$ are assumed to be constant values.\vspace{-0.35cm}

\subsection{Simulation Results on the ACSim Model}
\begin{figure*}[h!]
	\begin{center}
		\includegraphics[scale=0.35]{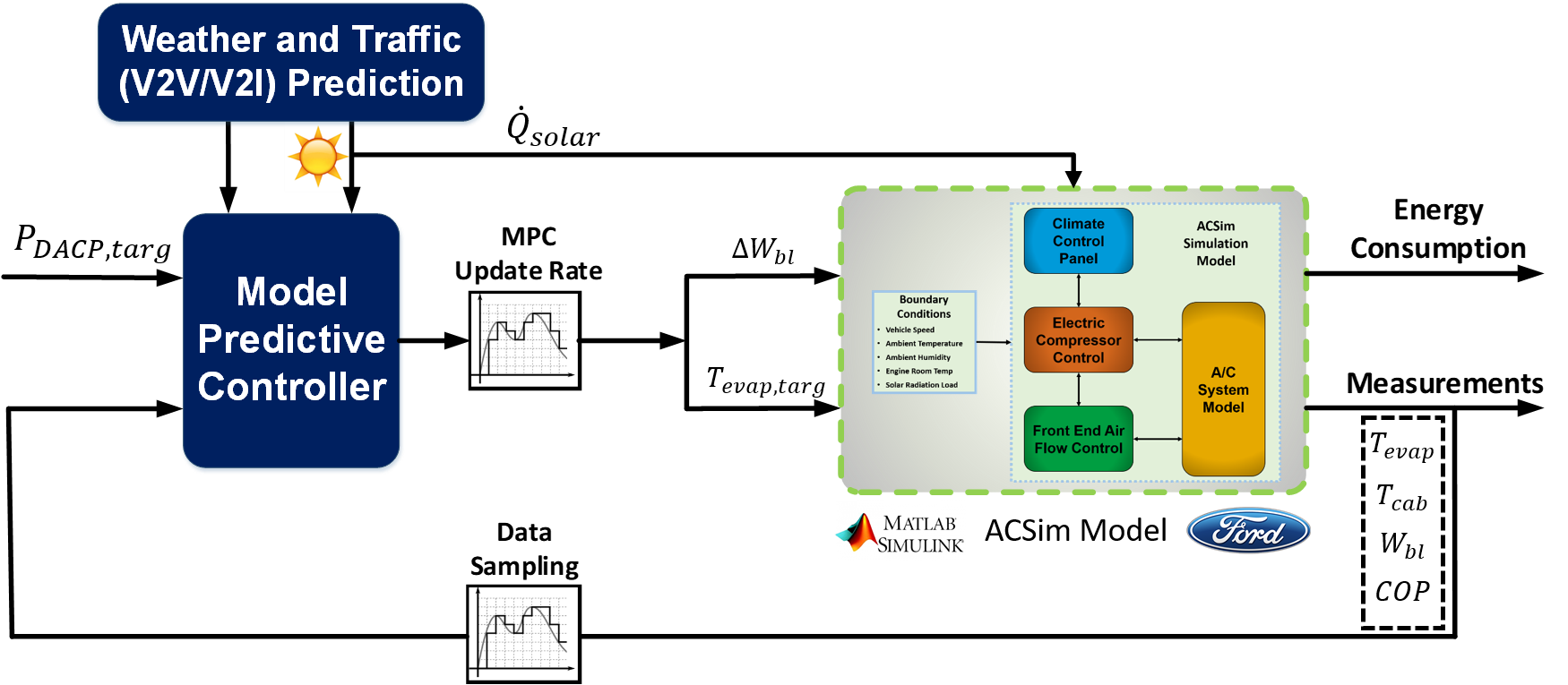}\vspace{-0.2cm}    
		\caption{Schematics of integrating the MPC-based PCS with ACSim model in Simulink\textsuperscript{\textregistered}.} 
		\label{fig:ACcontroldiagram}\vspace{-0.7cm}
	\end{center}
\end{figure*}

Next, the proposed control strategy is integrated in closed-loop with the ACSim model. Fig.~\ref{fig:ACcontroldiagram} illustrates the implementation  in Simulink\textsuperscript{\textregistered}. The model predictive controller takes sensor measurements, predefined $P_{DACP,targ}$ trajectories, and future vehicle speed from the traffic prediction as inputs, solves the optimization problem defined by (\ref{eqn:MPC-PCS}), and provides the control inputs to the ACSim model. 

\begin{figure}[h!]
	\begin{center}
		\includegraphics[width=4.0cm]{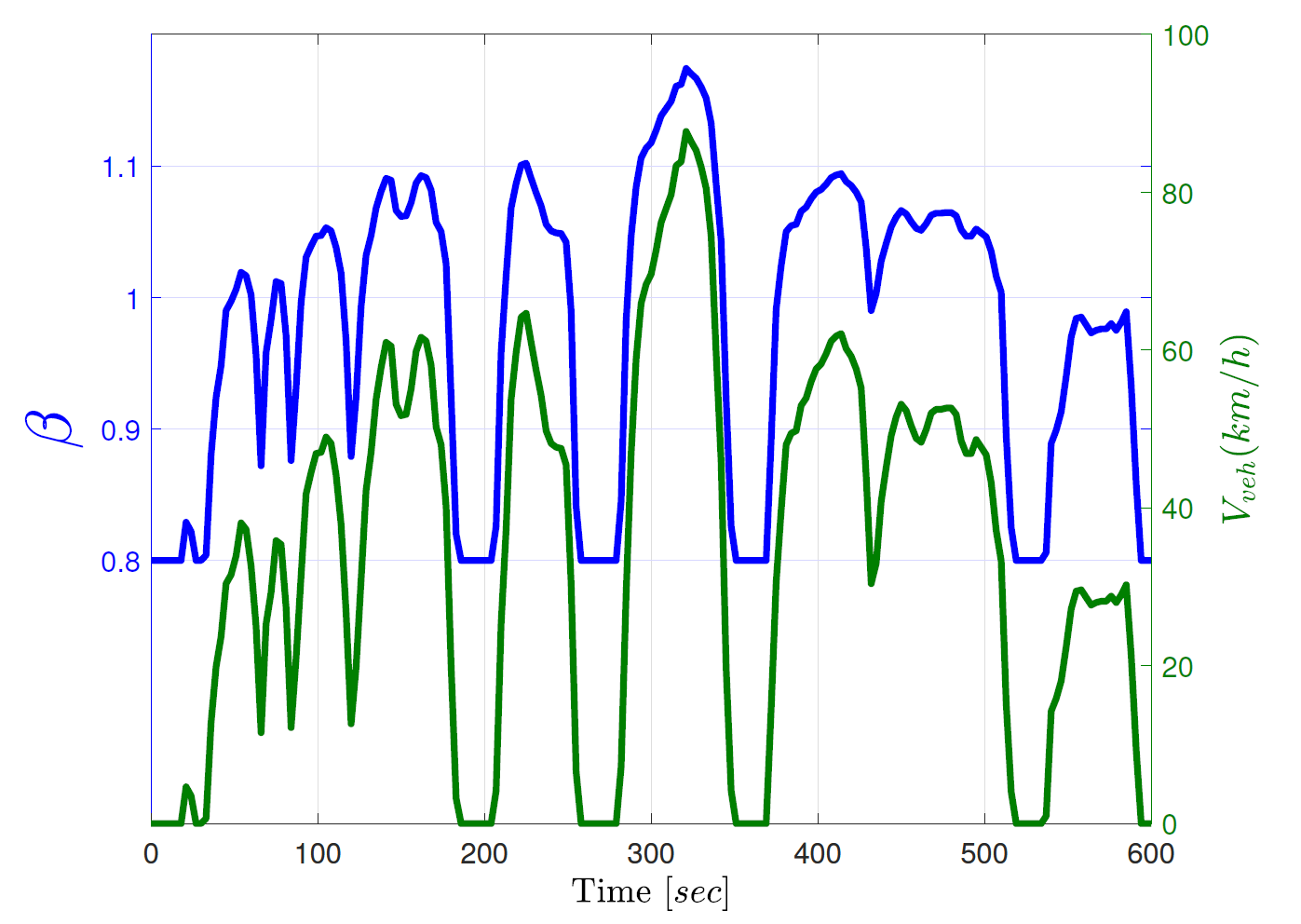} \vspace{-0.4cm}
		\caption{Heuristic design of speed-dependent $\beta$.} 
		\label{fig:Spd_denpendent_beta} 
	\end{center}\vspace{-0.6cm}
\end{figure}

In this case study, the proposed MPC controller updates the control inputs every $3~sec$, while the outputs from ACSim model is originally sampled at $0.1~sec$. The same cabin cool-down process is considered and $P_{DACP,targ}$ trajectory is calculated from a Ford benchmark case over SC03 cycle. In addition, a heuristic design of $\beta$ with respect to the speed profile from SC03 cycle is applied. The dependence of $\beta$ on different vehicle speed can be seen from Fig.~\ref{fig:Spd_denpendent_beta}. The idea behind such heuristic design of $\beta$ coincides with the exploration of the speed sensitivity of A/C operation, which is that energy efficiency may be improved by shifting the A/C load from low efficiency region (at low vehicle speed) to high efficiency region (at high vehicle speed). In the simulation with time-varying $\beta$, the vehicle speed over the prediction horizon is assumed to be known via connectivity technology, thus the values of $\beta$ over the prediction horizon are also available.

\begin{figure}[h!]
	\begin{center}
		\includegraphics[width=8cm]{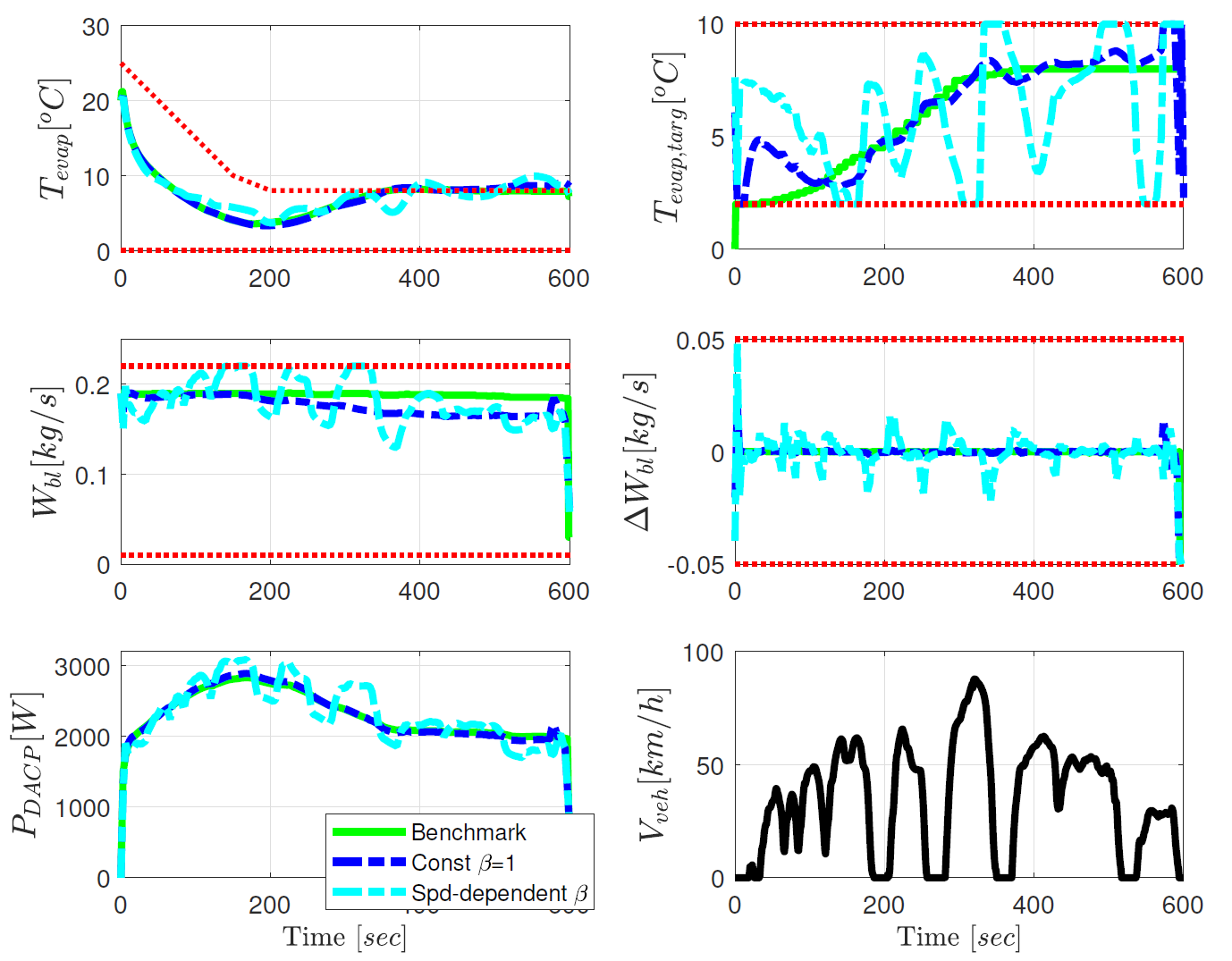} \vspace{-0.45cm}
		\caption{Comparison between the proposed PCS and the benchmark case on the ACSim model (key control variables).} \vspace{-0.65cm}
		\label{fig:MPC_ACSIM} 
	\end{center}
\end{figure}

\begin{figure}[h!]
	\begin{center}
		\includegraphics[width=7cm]{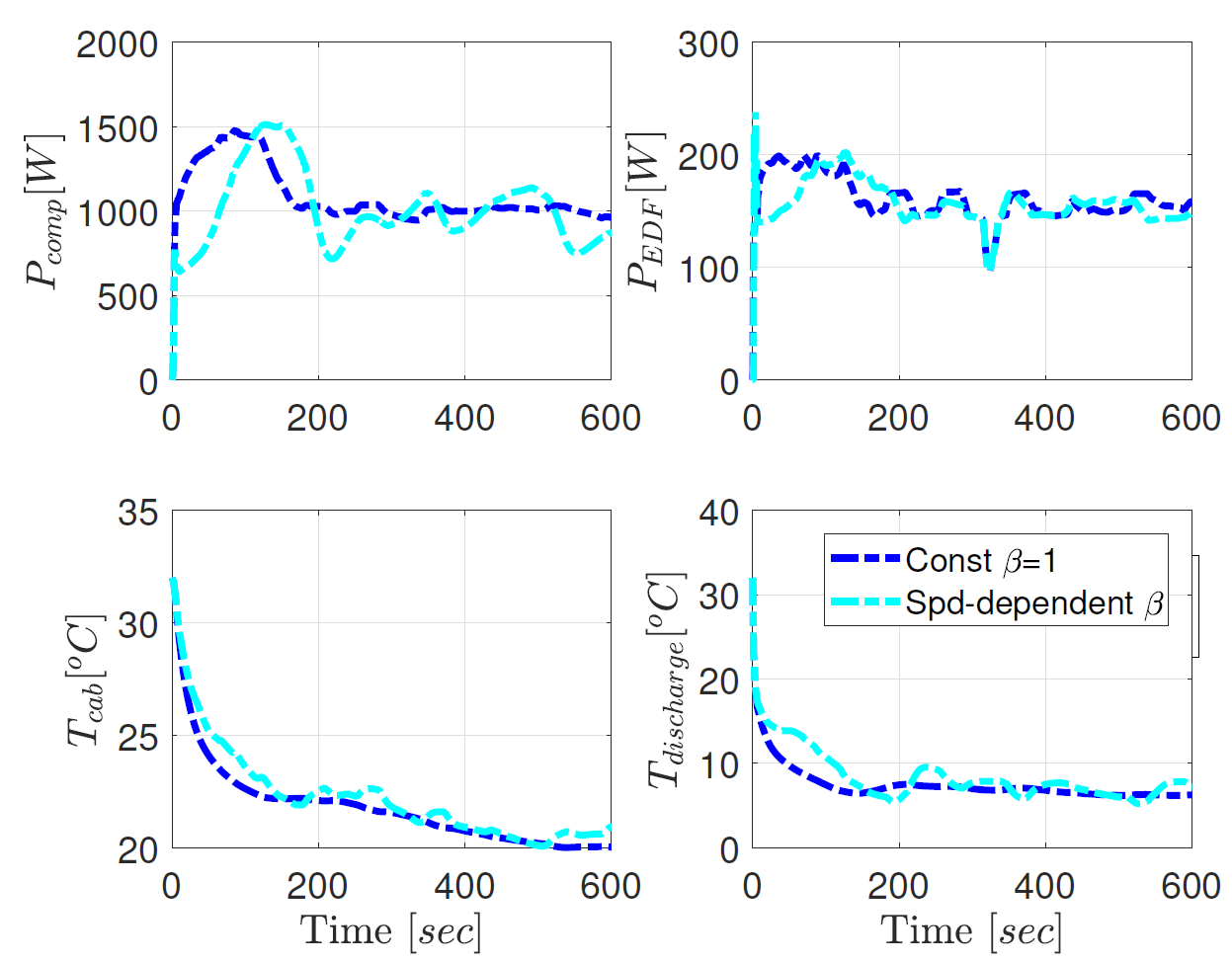} \vspace{-0.4cm}
		\caption{Comparison between the proposed PCS and the benchmark case on the ACSim model (A/C energy consumptions and temperatures) .} 
		\label{fig:MPC_ACSIM_energy} 
	\end{center}\vspace{-0.5cm}
\end{figure}

\begin{figure}[h!]
	\begin{center}
		\includegraphics[width=5cm]{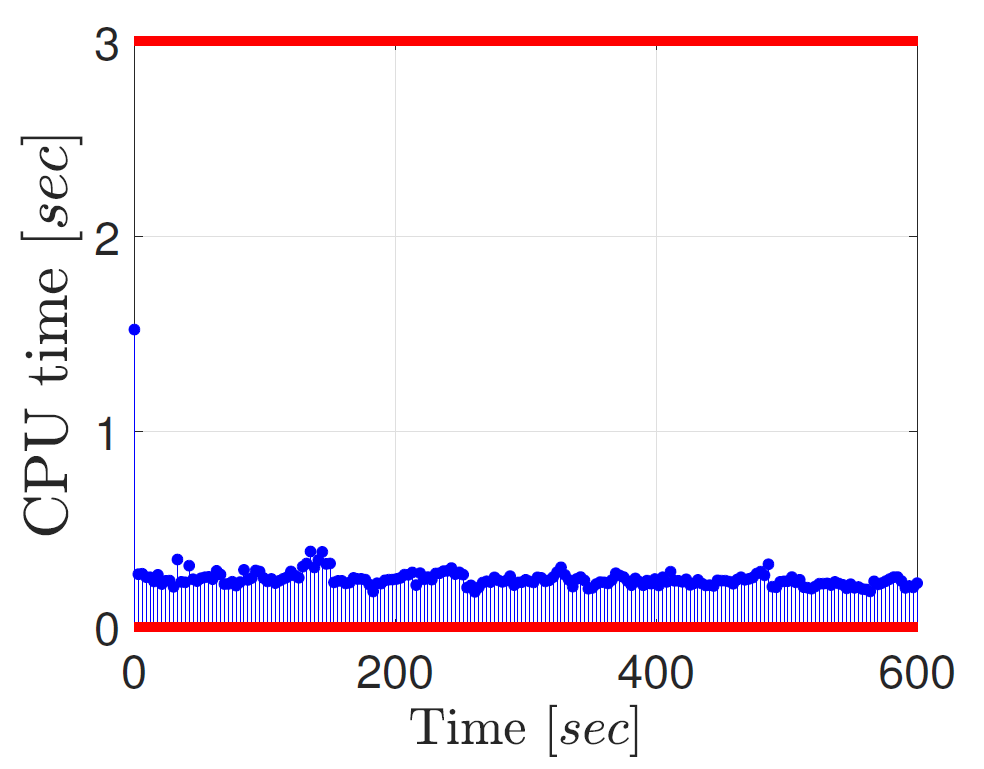} \vspace{-0.4cm}
		\caption{Elapsed CPU time for computing MPC solution for each control instant on ACSim model.} 
		\label{fig:CPUtime} 
	\end{center}\vspace{-0.8cm}
\end{figure}

Fig.~\ref{fig:MPC_ACSIM} compares the benchmark case with the NMPC results with constant $\beta$ and speed-dependent $\beta$, respectively. As observed from the results, for the constant $\beta$ case, the NMPC regulates the control inputs to achieve precise tracking of the $P_{DACP,targ}$ trajectories. For the speed-dependent $\beta$, the actual $P_{DACP}$ varies around the target. In addition, clear coordination between the control inputs and vehicle speed can be seen in the speed-dependent $\beta$ case indicating successful load shift as intended. In this simulation, $T_{cab}$ and $COP$ are assumed to have constant values along prediction horizon for each control iteration and are updated using measurements at every sampling instant. Additional system responses including the trajectories of $P_{comp}$, $P_{EDF}$, $T_{cab}$ and $T_{discharge}$ are shown in Fig.~\ref{fig:MPC_ACSIM_energy}. Detailed energy consumptions of different cases are reported in Table~\ref{tb:Energy}. It can be seen that, compared with the benchmark case, the total A/C energy consumption is reduced by $4.9\%$ for the MPC results with constant $\beta$. This is because for matching the $P_{DACP,targ}$, the MPC-based controller tends to reduce the air flow ($W_{bl}$) towards the end of the cycle, which results in the same pull-down period of the cabin temperature ($T_{cab}$) but slight increase in final cabin temperature (with difference less than $1 ^oC$). In other words, the actual cooling capacity of the A/C system is reduced for the MPC case while achieving the same occupant thermal comfort level according to the proposed metric. If we compare the MPC results with speed-dependent $\beta$ with the ones with constant $\beta$, we can see that the energy consumption of the A/C system may be further reduced by $0.8\%$ while providing $1.1\%$ higher $E_{DACE}$. The energy saving achieved by A/C load shifting can be even higher if designing $\beta$ optimally instead of designing it in a heuristic way. Fig.~\ref{fig:CPUtime} reports the elapsed CPU time for each control iteration compared with $3~sec$ for the MPC sampling time. This result is obtained based on a $2.9~GHz$ Windows computer for the speed-dependent $\beta$ case considered in this section. Note that the worst case execution time is significantly lower than the available time. These results suggest that our NMPC approach could be computationally feasible even in slower ECU as the ECU implementation will be based on highly optimized C-code (rather than Matlab) that, based on our past experience, is likely to offset the processor differences. 

\begin{table*}[t!]
	\vspace{+0.7cm}
	\caption{A/C system energy consumption comparisons of applying constant $\beta=1$ and speed dependent $\beta$ with respect to the benchmark.} \vspace{-0.4cm}
	\begin{center}
		\label{tb:Energy}
	\begin{tabular}{|c|c|c|c|c|}
		\hline
		& $E_{DACE}$ {[}kJ{]} & $E_{comp}$ {[}kJ{]} & $E_{EDF}$ {[}kJ{]} & $E_{tot}$ {[}kJ{]} \\ \hline
		Benchmark    & 1378.4              & 689.4               & 103.2               & 792.6              \\ \hline
		Constant $\beta=1$    & 1377.7 (-0.1\%)             & 653.4               & 100.6               & 754.0 (-4.9\%)              \\ \hline
		Spd-dependent $\beta$ & 1392.0 (+1.0\%)     & 647.2      & 100.6      & 747.8 (-5.7\%)     \\ \hline
	\end{tabular}\vspace{-0.55cm}
    \end{center}
\end{table*}

\section{CONCLUSIONS}\label{sec:5}
A novel MPC-based precision cooling strategy (PCS) was proposed in this paper to exploit the energy saving opportunities for the thermal management of automotive A/C system. The proposed PCS was designed to precisely track the prescribed and time-varying trajectory of the discharge air cooling power (DACP), which represents the desired cabin cooling requirement. A physics-based A/C system model, ACSim, developed by Ford Motor Company was adopted as the virtual test bench in this study. To satisfy the requirements of MPC-based design, a simplified predictive model was developed based on the ACSim model responses. Next, the proposed MPC-based PCS was formulated by solving a nonlinear optimization problem, which minimizes (i) the tracking error with respect to the DACP trajectory target, (ii) the energy consumption of the A/C system. The performance of the proposed PCS was evaluated in closed-loop with ACSim model. Accurate tracking performance and constraint enforcement on system states and inputs have been demonstrated. It was also shown that, comparing with a Ford benchmark case over SC03 test cycle, the MPC-based solution of tracking the same DACP trajectory saves $4.9\%$ electrical energy at the expense of slightly increased cabin temperature towards the end of the simulation. In addition, by exploiting the vehicle speed preview, the A/C system energy consumption can be further reduced. Future work will target computing the $P_{DACP,targ}$ based on detailed passenger thermal comfort model as well as designing an optimal load shifting strategy and validating the proposed PCS in vehicle experiments.  \vspace{-0.2cm}

\section{acknowledgment} The authors would like to thank Christopher Greiner from Ford Motor Company for valuable discussions and comments on this work. \vspace{-0.2cm}





\begin{thebibliography}{99}
\bibitem{Zhang2011} 
B. Zhang, M. Chris, and M. Zhang, ``Charge-depleting control strategies and fuel optimization of blended-mode plug-in hybrid electric vehicles," {\em IEEE Transactions on Vehicular Technology}, vol. 60, no. 4, pp. 1516–1525, 2011.

\bibitem{Zhang2012} 
M. Zhang, Y. Yan, and M. Chris, ``Analytical approach for the power management of blended-mode plug-in hybrid electric vehicles," {\em IEEE Transactions on Vehicular Technology}, vol. 61, no. 4, pp. 1554–1566, 2012.

\bibitem{Guanetti18}
J. Guanetti, Y. Kim, and F. Borrelli, ``Control of connected and automated
vehicles: state of the art and future challenges,'' {\em Annual Reviews
in Control}, vol. 45, p18-40, 2018.

\bibitem{Amey2016}
A. Karnik, A. Fuxman, P. Bonkoski, M. Jankovic, and J. Pekar, ``Vehicle powertrain thermal management system using model predictive control,'' {\em SAE International Journal of Materials and Manufacturing}, vol. 9, no. 3, pp. 525-533, 2016.

\bibitem{Feru2016}
E. Feru, N. Murgovski, B. Jagar, and F. Willems, ``Supervisory control of a heavy-duty diesel engine with an electrified waste heat recovery system,'' {\em Control Engineering Practice}, vol. 54, pp. 190-201, 2016.

\bibitem{Xia2017}
G. Xia, L. Cao, and G. Bi, ``A review on battery thermal management in electric vehicle application,” {\em Journal of Power Sources}, vol. 367, pp. 90-105, 2017.

\bibitem{Snyder2014}
K. Snyder, and J. Ku, ``Optimization for plug-in vehicles-waste heat recovery from the electric traction motor,'' (No. 2014-01-1921) {\em SAE Technical Paper}, 2014.

\bibitem{Rugh2008} 
J. Rugh, and R. Farrington, ``Vehicle ancillary load reduction project close-out report: an overview of the task and a compilation of the research results," {\em Technical Report, NREL/TP-540-42454}, 2008.

\bibitem{Jeffers2016} 
M. Jeffers, L. Chaney, and J. Rugh, `` Climate control load reduction strategies for electric drive vehicles in cold weather," {\em SAE International Journal of Passenger Cars -- Mechanical Systems}, Vol. 9, pp.75-82, 2016.

\bibitem{Rask2014}
E. Rask, ``Ford focus BEV in-depth (Level 2) testing and
analysis," {\em Presented at Vehicle Systems Analysis Technical
Team (VSATT) meeting}, Apr. 2014.

\bibitem{Jeffers2015} 
M. Jeffers, L. Chaney, and J. Rugh, ``Climate control load reduction strategies for electric drive vehicles in warm weather," (No. 2015-01-0355) {\em SAE Technical Paper}, 2015.


\bibitem{Zhang2017}
Q. Zhang, Y. Meng, C. Greiner, C. Soto, W. Schwartz, and M. Jennings, ``Air conditioning system performance and vehicle fuel economy trade-offs for a hybrid electric vehicle," {\em SAE Technical Paper}, 2017-01-0171, 2017. 

\bibitem{Hao2018}
H. Wang, I. Kolmanovsky, M. Amini, and J. Sun, ``Model predictive
climate control of connected and automated vehicles for improved
energy efficiency,'' {\em in American Control Conference}, June 27-29, 2018, Milwaukee, WI, USA.

\bibitem{Reza18}
M. Amini, H. Wang, X. Gong, D. Liao-McPherson, I. Kolmanovsky, and J. Sun, ``Cabin and battery thermal management of connected and automated HEVs for improved energy efficiency using hierarchical model predictive control," submitted to {\em IEEE Transactions on Control Systems Technology}.

\bibitem{Kelman11} 
A. Kelman and F. Borrelli, ``Bilinear model predictive control of a HVAC system using sequential quadratic programming,'' {\em Proc. of the 18th IFAC World Congress,} Milano, Italy, 2011.

\bibitem{Oldewurtel12}
F. Oldewurtel, A. Parisio, C. N. Jones, D. Gyalistras, M. Gwerder, V. Stauch, B. Lehmann, and Manfred Morari, `` Use of model predictive control and weather forecasts for energy efficient building climate control,'' {\em Energy and Buildings}, Vol. 45, pp. 15-17, 2012.

\bibitem{Reza19_2}
M. Amini, X. Gong, Y. Feng, H. Wang, I. Kolmanovsky, and J. Sun, ``Sequential optimization of speed, thermal load, and power split in connected HEVs," in {\em American Control Conference}, July 10-12, 2019, Philadelphia, PA, USA.

\bibitem{Zhang16}
Q. Zhang, S.E. Li, and K. Deng, ``Automotive air conditioning: optimization, control and diagnosis,'' Springer, 2016.





\bibitem{Bhatti97}
M.S., Bhatti, ``A critical look at R-744 and R-134a mobile air conditioning systems,'' {\em SAE Technical Paper}, 970527, 1997. 


\bibitem{mpctools}
M.J. Risbeck, and J.B. Rawlings, ``MPCTools: nonlinear model predictive control tools for CasADi,'' 2016.

\bibitem{CAS18}
J. A. Andersson, J. Gillis, G. Horn, J. B. Rawlings, and M. Diehl, ``CasADi -- A software framework for nonlinear optimization and optimal control,"
{\em Mathematical Programming Computation}, in press, 2018.

\end{thebibliography}
\end{document}